\documentclass[sigconf]{acmart}
\usepackage{xcolor,soul}
\AtBeginDocument{%
  }

\copyrightyear{2024}
\acmYear{2024}
\setcopyright{rightsretained}
\acmConference[UbiComp Companion '24]{Companion of the 2024 ACM International Joint Conference on Pervasive and Ubiquitous Computing Pervasive and Ubiquitous Computing}{October 5--9, 2024}{Melbourne, VIC, Australia}
\acmBooktitle{Companion of the 2024 ACM International Joint Conference on Pervasive and Ubiquitous Computing Pervasive and Ubiquitous Computing (UbiComp Companion '24), October 5--9, 2024, Melbourne, VIC, Australia}
\acmDOI{10.1145/3675094.3677547}
\acmISBN{979-8-4007-1058-2/24/10}

\acmSubmissionID{ubide1153}

\begin{document}
\title[Seamless Detection of Time-Killing Using Continuous Screen Text and On-Device LLMs]{ScreenTK: Seamless Detection of Time-Killing Moments Using Continuous Mobile Screen Text and On-Device LLMs}
\author{Le Fang}
\affiliation{%
 \institution{The University of Melbourne}
 \city{Melbourne}
  \country{Australia}
    }
  \email{le.fang2@unimelb.edu.au}

\author{Shiquan Zhang}
\affiliation{%
  \institution{The University of Melbourne}
  \city{Melbourne}
  \country{Australia}}
\email{shiquan.zhang@student.unimelb.edu.au}

\author{Hong Jia}
\affiliation{%
 \institution{The University of Melbourne}
 \city{Melbourne}
  \country{Australia}
    }
\email{hong.jia@unimelb.edu.au}

\author{Jorge Goncalves}
\affiliation{%
 \institution{The University of Melbourne}
 \city{Melbourne}
  \country{Australia}
    }
\email{jorge.goncalves@unimelb.edu.au}

\author{Vassilis Kostakos}
\affiliation{%
 \institution{The University of Melbourne}
 \city{Melbourne}
  \country{Australia}
  }
\email{vassilis.kostakos@unimelb.edu.au}

\renewcommand{\shortauthors}{Le Fang, Shiquan Zhang, Hong Jia, Jorge Goncalves, \& Vassilis Kostakos}

%---------------------------------------
\begin{abstract}
% Smartphones are inseparable for people's digital life, connecting users to the world with the endless information flow. Unavoidably, this constant stream might be overwhelming, underscoring the necessity of developing methods for delivering high-stakes notifications that reducing interruption while increasing engagement. One effective approach is to detect moments when users are most receptive to incoming messages, known as "attention surplus" moments. Within this framework, "time-killing" moments have been considered as a specific type of attention surplus. The existing method for "time-killing" detection is using screenshots taken every 5 seconds, ignoring phone usage between the 5-second gaps. In this paper, we showed that 50\% of time-killing instances cannot be detected by using the screenshots, proposing a novel method that seamlessly detects time-killing by continuous screen text monitoring on the smartphones. Our experiment results suggested that screen text could capture comprehensive information about the time-killing behavior. The richness of the screen text was also illustrated by showing that the large language models could generate detailed summary of users' time-killing behavior. Hence, we proposed that screen text is effective for detecting the time-killing, improving self-awareness of the digital life, and empowering the self-control over users' smartphone usage.

Smartphones have become essential to people’s digital lives, providing a continuous stream of information and connectivity. However, this constant flow often depletes users' limited attentional resources and time, leading to decreased productivity and increased stress levels. This issue underscores the need for tools that empowers users to maximize their potential for achieving personal objectives. One effective approach is to identify ``time-killing'' moments---a specific type of attention surplus---during which users seek to fill perceived free time without a specific purpose.
Recent work has utilized screenshots taken every 5 seconds to detect time-killing activities on smartphones. However, this method often misses to capture phone usage between intervals. We demonstrate that up to 50\% of time-killing instances go undetected using screenshots, leading to substantial gaps in understanding user behavior. To address this limitation, we propose a method called ScreenTK that detects time-killing moments by leveraging continuous screen text monitoring and on-device large language models (LLMs). Screen text contains more comprehensive information than screenshots and allows LLMs to summarize detailed phone usage. To verify our framework, we conducted experiments with six participants, capturing 1,034 records of different time-killing moments. Initial results show that our framework outperforms state-of-the-art solutions by 38\% in our case study. 

\end{abstract}


\maketitle

\section{Introduction}
\label{introduction}

\begin{figure*}[h]
    % \vspace{-3em}

    \centering
    \includegraphics[width=1\textwidth]{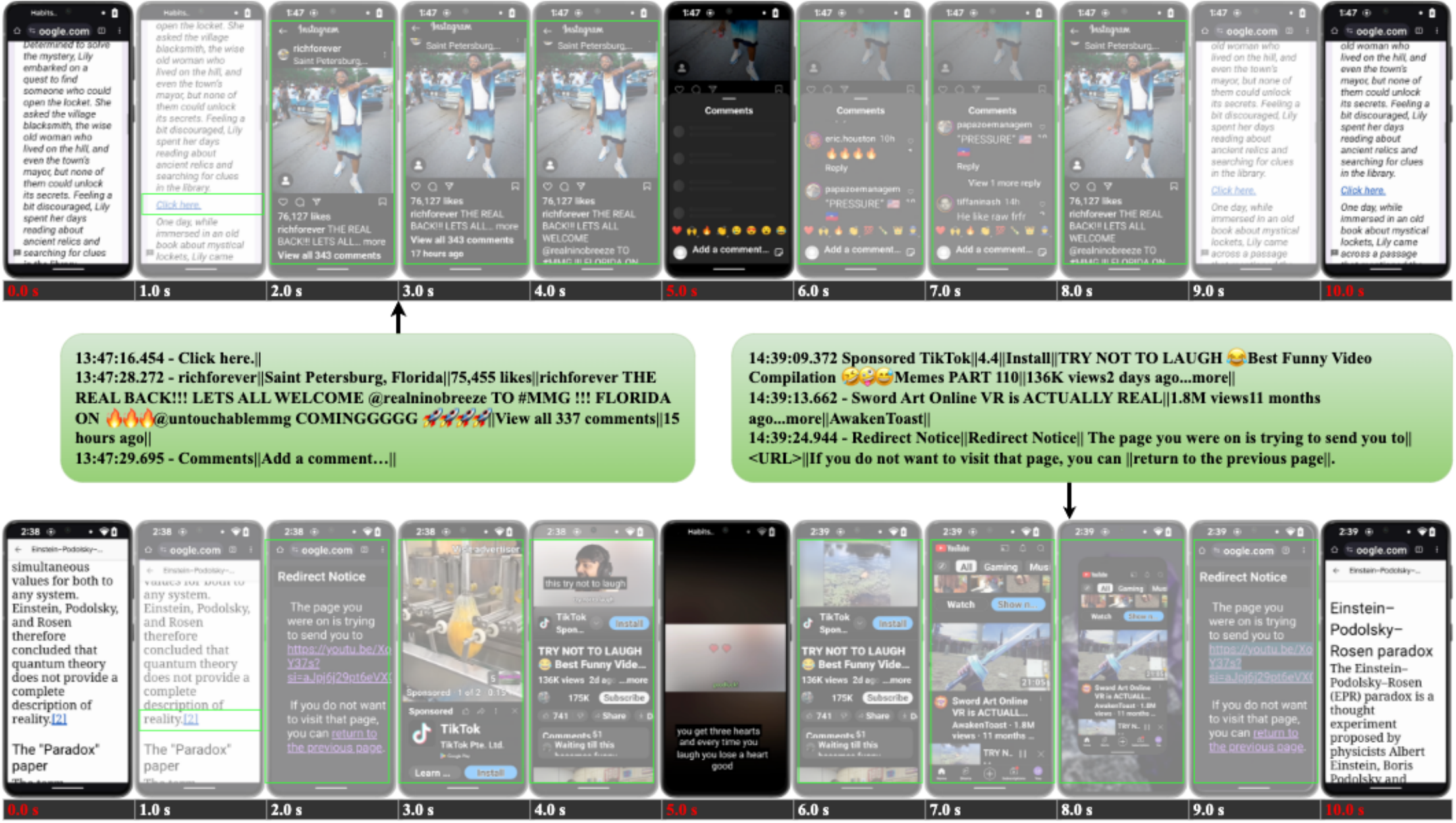} %width=10cm
    \caption{Comparison between ScreenTK and SOTA screenshot-based method~\cite{10.1145/3544548.3580689} for detecting time-killing moments. The green boxes highlight the records of time-killing instances captured by the proposed framework, ScreenTK. Top: explicit time-killing study. Bottom: implicit time-killing study. Transparent: ScreenTK only. Non-Transparent: Both ScreenTK and Screenshot.}
    \label{fig:gap1}
\end{figure*}

% (up.. passive) 
% Smartphones are essential for people's digital life. Users received numerous information from their phones. It is necessary to find a way that delivers important notifications to users with less distraction but more engagement. The main method to find such way is to predict moments that users are receptive to delivered contents, as known as moments of "attention surplus" \citep{pielot2015attention}. \citet{10.1145/3544548.3580689} found time-killing moments are one type of "attention surplus" that offers a suitable chance to delivering contents to users because users had no specific goal and wanted to fill their perceived free time during time-killing \citep{lukoff2018makes, oulasvirta2012habits}, e.g., waiting for a train or during an unattractive speech \citep{isaacs2009mobile}.
% \hj{
Smartphones are essential to modern digital life, providing users with a constant influx of information. %However, this constant stream can sometimes be overwhelming, causing users to simply pass time instead of engaging in meaningful activities. 
However, this nonstop stream can sometimes be overwhelming, depleting users' limited attentional resources and hindering the pursuit of personal goals (e.g., learning a language and reading a book). This challenge highlights the need for effective tools that offer users the choice to detect opportune moments to maximize their potential for achieving these objectives.
One promising approach is to identify moments when users are most receptive to incoming content, known as "attention surplus" 
moments~\citep{pielot2015attention}. Within this framework, ``time-killing'' moments have been identified as a specific type of attention surplus~\citep{10.1145/3544548.3580689}. Specifically, during ``time-killing'' moments, users, having no specific goal, seek to fill their perceived free time \citep{lukoff2018makes, oulasvirta2012habits}, such as when they are waiting for a train or listening to an unengaging speech \citep{isaacs2009mobile}.
% }
% \hj{wait for train is time killing?}

% As such, capturing time-killing moments is beneficial for designing effective notification system on smartphones, allowing people to engage in more productive tasks when utilizing the free time \citep{pielot2017beyond}. Additionally, time-killing detection could inform researchers about duration and distribution of users' daily time-killing behaviors \citep{10.1145/3544548.3580689}. \hj{← These still discuss the background which should finish in the first paragraph. Delete and go straight to discuss the related work and the gap: What leads to the gap.} \citet{10.1145/3544548.3580689} introduced a deep-learning-based fusion model for time-killing moment detection on smartphones using screenshots taken every 5 seconds and the corresponding phone sensor data. However, this method for time-killing detection has a serious limitation because many time-killing instances that lasted less than 5 seconds were not captured by the screenshots. To address this issue, we proposed a seamless approach to detect time-killing moments by using continuous screen text collected by a digital phenotype tool AWARE-Light \citep{10.1145/3613904.3642347}.

To capture moments of distraction, existing works focus on utilizing screenshots~\cite{10.1145/3544548.3580689} as the main information source and traditional machine learning algorithms for modeling. Specifically, a state-of-the-art (SOTA) method~\cite{10.1145/3544548.3580689} employs a 30-second duration with 5-second interval screenshots to determine whether a user is distracted, using a CNN-LSTM structure to train the model in a supervised manner. However, the 5-second duration can miss significant phone usage information. For instance, detection may fail if users periodically switch to social media apps during the 5-second intervals. Additionally, supervised models are not adept at summarizing and generating useful information to enhance users' self-awareness of their phone usage. These limitations highlight the need for more robust and effective methods that can better capture and analyze user's phone usage.
% }

In this paper, we propose ScreenTK, a novel framework to seamlessly capture "time-killing" moments using continuous screen text monitoring and large language models (LLMs). Specifically, we propose using screen text to collect distraction moments as it provides more comprehensive information to capture the user's phone usage compared to screenshots. We then apply SOTA LLMs to identify these moments and summarize key information, such as preferences, wish lists, and to-do lists, offering the user a more fine-grained understanding of their daily phone usage. 

To evaluate the proposed framework, we designed three case studies involving six participants and captured 1,034 records containing time-killing moments. Compared with SOTA screenshot baselines, we observed that the proposed method significantly outperformed them by 38\% in our case study. We envision that the proposed framework can significantly help users in shaping their self-awareness of daily phone usage. 
% }
% We evaluated and compared the performance of screen text and screenshots on the time-killing detection in a trial experiment. The result suggested that the screen text could obtain more fine-grained information about time-killing behaviors than the screenshots. In addition, we designed prompt for zero-shot learning with large language models (LLMs) to analyze screen text of time-killing behavior. The LLMs' output demonstrated that screen text contained in-depth time-killing information. Additionally, screen text costs much less than screenshots in terms of storage and processing. In the future, we plan to explore screen text of time-killing behavior with on-device learning to address privacy and security concerns caused by LLMs.

% The paper is organized as follows: Section \ref{study_design} introduces the motivation and study design of time-killing detection. Next, the experimental method is presented in Section \ref{method}, followed by the evaluation results in Section \ref{evaluation}. The paper continues with related works in Section \ref{related_works} and concludes in Section \ref{conclusion}.
\section{Related Works}
\label{related_works}
This section summarizes some existing works on phone usage behaviors in Section \ref{phone_usage_research} and the time-killing detection in Section \ref{time_killing_detection}.
% \vspace{-1em}

% \vspace{-2.5em}
\subsection{Phone Usage Behaviors}
\label{phone_usage_research}
Self-report methods (e.g., via interviews and diaries \citep{10.1145/587078.587092}) were often used in early phone usage studies to help users understand their usage patterns, motivations, and behaviors, but these methods were criticized due to inaccurate or biased predictions \citep{10.1145/2070481.2070550, 10.1145/1247660.1247670}. In comparison, quantitative analysis of phone-usage logs has become more popular \citep{10.1145/1814433.1814453, 10.1145/2068816.2068847, fea2fb2aec0b47c78dd185d58e36c382}. Leveraging large-scale datasets collected by mobile sensors, many researchers have focused on modeling phone-use behavior, such as predicting smartphone screen use \citep{10.1145/2971648.2971669} and classifying app usage by combining phone logs with experience-sampling method data \citep{10.1145/3229434.3229441}. However, phone logs of system data (e.g., screen events and app states) are limited in capturing the complexity of smartphone use. 

To obtain a more comprehensive picture of people's digital life and behaviors on smartphones, previous work has explored using screenshots to analyze app usage \citep{10.1145/2628363.2628377, abc69ccf1c3048e0926491266c79c995}. However, while screenshots can partly reveal users' actions, they are not continuous, and phone use information could be missed during the time gaps. In comparison, we leverage continuous screen text as the information source and utilize LLMs to seamlessly help users understand their phone usage.

\subsection{Time-Killing Detection}
\label{time_killing_detection}
\citet{10.1145/3544548.3580689} stated that time-killing on smartphones is ubiquitous and offers opportunities to deliver content to users. To study time-killing behavior, the authors developed an Android app called Killing Time Labeling (KTL) to collect and annotate screenshots and phone-sensor data (e.g., Android accessibility events, screen status, and type of transportation) of users' app usage. The app runs as a background service that automatically takes screenshots every 5 seconds (only when the screen is on). However, as mentioned in Section \ref{phone_usage_research}, the screenshots might miss many details of time-killing moments that happen between the 5-second gaps. In comparison, we propose a novel approach that uses a digital phenotype tool, AWARE-Light \citep{10.1145/3613904.3642347}, to collect screen text of app usage to continuously monitor time-killing behaviors on smartphones.

% Understanding smartphone usage is one of the main topics for digital wellbeing study. However, limited work discussed
% \citet{10.1145/3544548.3580729} conducted a systematic review on attention capture damaging patterns (ACDPs) and initiated typology of the patterns (e.g., infinite scrolling). The authors define ACDP as:

% \begin{quote}
%     A recurring pattern in digital interfaces that a designer uses to exploit psychological vulnerabilities and capture attention, often leading the user to lose track of their goals, lose their sense of time and control, and later feel regret.
% \end{quote}

% %how to measure endless scrolling? 
% %what are the research questions?
% %features analysis
% %LLM?
% %Use LLM to get limitations of the related works

% \citet{10.1145/3544548.3580729} summarized 2 criteria in mechanisms and 3 in impacts that determine ACDPs. For mechanisms, ACDPs utilize 1) psychological weaknesses and 2) automate the user experience. For impacts, ACDPs cause users to 3) forget their goals, 4) lose sense of time and agency, and 5) users feel disappointed after using a digital service implementing ACDPs. The authors call attention to users' digital wellbeing for new design processes. They stated that avoiding ACDPs might initially reduce lower user engagement, but alternative designs would increase the long-term user loyalty.
 
\section{Method}
\label{method}
This section will first discuss how we (1) seamlessly capture screen text information ( Section \ref{screen_text}) and (2) detect time-killing moments via on-device LLMs (Section \ref{on-device}).

% Section \ref{screen_text} introduces the screen text sensor of AWARE-Light and the data export procedure. Section \ref{screenshots} discusses the extraction of screenshots for time-killing moments.

\subsection{Seamless Capture of Screen Information}
\label{screen_text}
To capture continuous screen information, we leverage the AWARE-Light app to extract phone usage information. Specifically, AWARE-Light is based on the Android Accessibility API, enabling the collection of various screen usage information including screen status (on/off and unlocked/locked), screen text, and touch events (click and scroll). In this study, we focus solely on screen text information. In detail, we install and configure the AWARE-Light app on a Google Pixel 8 to capture the screen information. After that, we extract each participant's phone usage information into a CSV file on the phone and feed it into an LLM for time-killing detection.
% \begin{figure}[h]
%     \centering
%     \includegraphics[width=0.5\textwidth]{figs/aware-light.png} %width=10cm
%     \caption{Main user interface of AWARE-Light.}
%     \label{fig:gap}
% \end{figure}

\begin{figure}[t]
    \centering
    \includegraphics[width=1\columnwidth]{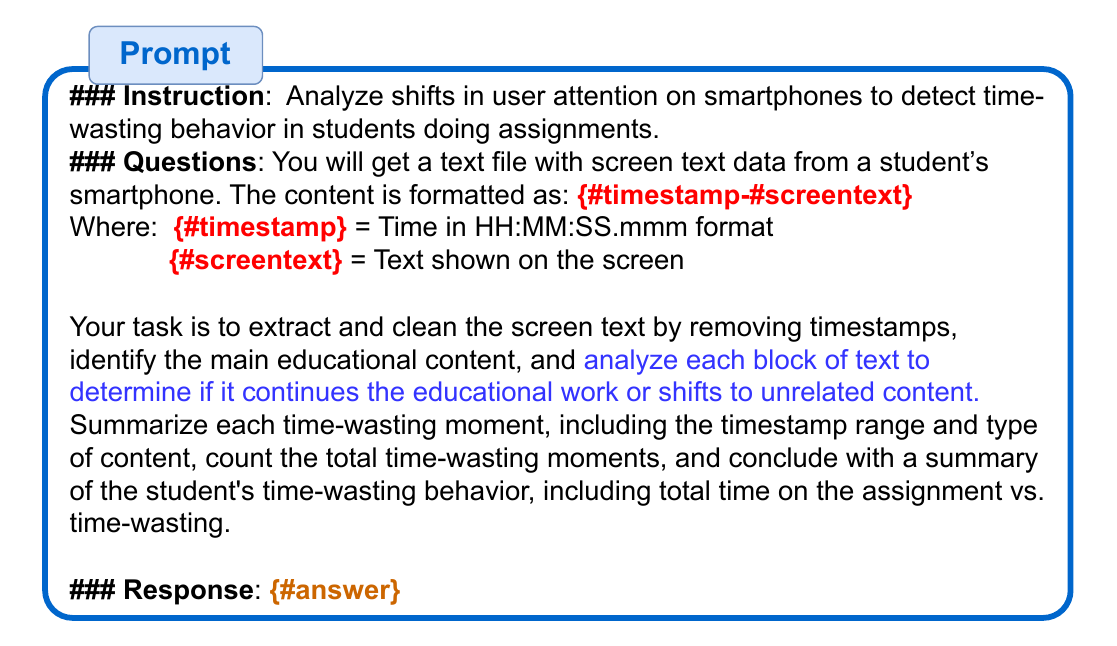} %width=10cm
    \caption{Prompt engineering for analyzing smartphone screen text data to understand user time-killing behavior, including data format, expected insights, and response structure.}
    \label{fig:gap}
\end{figure}

\subsection{Time-killing Detection via On-device LLMs}\label{on-device}
Screen text usually contains significant amount of data (usually millions of tokens), make traditional machine learning models such as SOTA time-killing used CNN and LSTM incapable to handle. In comparison, recent LLMs are capable to throughput millions of token size information. Also, LLMs are inherently well-suited for understanding and analyzing text-based information due to their extensive pretraining on natural languages. As such, it is reasonable to choose LLMs to capture user's time-killing moments. To achieve this and protect user's privacy, we utilize an on-device open sourced LLMs model called LLama3~\footnote{https://github.com/meta-llama/llama3} deployed directly on smartphone.

% To help LLMs understand the screen text information, we propose a prompt engineering to analyze human time-killing behavior on smartphones using screen text data. The prompt will instruct the model to examine a CSV file containing two columns: {$\#timestamp$, $\#text$}. The {$\#timestamp$} column uses Unix timestamp format, while the {$\#text$} column displays the screen text at the given timestamp. The model will identify time-killing moments by detecting switches from reading textbooks to social media apps such as \textit{Facebook}, \textit{Instagram}, \textit{Netflix}, \textit{YouTube}, \textit{TikTok}, and many others. The analysis will involve counting instances of such switches and providing a step-by-step breakdown of the findings.
We design a prompt that involves analyzing shifts in user attention on smartphones to detect time-wasting behavior in students engaged in assignments for our case studies. Specifically, for the instructions, "You will get a text file with screen text data from a student’s smartphone" formatted as ${\#timestamp-\#screentext}$, where the timestamp is in ``\textit{HH:MM:SS.mmm}'' format and represents the time the screen text was captured. The task is to ``extract and clean the screen text by removing timestamps,'' identify the main educational content, and analyze each block of text to determine if it continues the educational work or shifts to unrelated content. The analysis (i.e., ${\#answer}$) will summarize each time-wasting moment, including the ``timestamp range and type of content,'' count the total number of time-wasting moments, and conclude with a summary of the student’s time-wasting behavior. This summary should include the ``total time on the assignment vs. time-wasting.'' This structured approach aims to provide a detailed understanding of how students manage their attention and the extent to which they are distracted by their smartphones during academic tasks.

\subsection{Case Studies}
\label{study_design}
% We proposed screen text as a more comprehensive data source to study time-killing moments. In other words, screen text could provide more fine-grained information about smartphone usage than screenshots. 
Our goal is to utilize screen text to improve people's self-awareness and empower self-control over their digital life on smartphones, ultimately benefiting users' well-being. To achieve this, two case studies were conducted to compare the performance of screen text and screenshots in detecting time-killing behaviors on smartphones. We recruited a total of six volunteer participants from our labs to join the experiment (different 3 participants for each case study). The explicit study aims to detect the time-killing moments that are intentionally triggered (Section \ref{explicit_study}) and the implicit study focuses on the time-killing moments that are spontaneous (Section \ref{implicit_study}). Data collection was conducted in accordance with ethics approval from our university.
% (3 of whom helped improve the study design)

\subsection{Explicit Time-killing Study}
\label{explicit_study}
In the explicit study, we aimed to proactively trigger time-killing moments for participants when they were engaged in a reading task. Specifically, a story of 516 words in English was assigned to three participants. After each paragraph, there was a URL embedded in a sentence that read "Click here." Additionally, there was a question about the content of the story. The question aimed to prevent participants from merely scanning the content, ensuring they paid more attention to reading the story. To enable a a fair comparison, we used 5-second intervals to capture screenshots and used AWARE-Light to capture screen text information for each user. Data was stored on the device automatically and fed into on-device LLMs when the study was finished.

% To verify the  Participants consented to record the phone's screen during the experiment. In Python scripts we extracted screenshots every 5 seconds using FFmpeg (a multimedia framework) and created a grid of the extracted screenshots using ImageMagick (an image manipulation tool) for each participant's screen recording.INSERT image of screenshots in a grid % INSERT image of screenshots in a grid

\subsection{Implicit Time-Killing Study}
\label{implicit_study}
In the implicit study, we aim to promote spontaneous time-killing behaviors occurring when participants are focusing on a reading task. Specifically, we designed a reading study using a scientific essay of 2,351 words. The URLs of popular internet memes were embedded in 20 citation brackets. Three participants joined this study without being notified of anything they should be aware of. Other configurations followed those used in the explicit time-killing text consistently.
\section{Results and Discussion}
\label{evaluation}
% \subsection{Baseline and Ground Truth}
% \label{screenshots}

% After excluding 2 participants who did not have time-killing instances, 
The explicit time-killing study collected 535 records and 11 minutes of screen text from another 3 participants (on average, 178 records and 4 minutes per participant). In comparison, the implicit time-killing study collected 499 records of screen text from three participants (on average, 11 minutes and 100 records per participant). The quantitative results are discussed in Section \ref{quantitative}, and the qualitative results are provided in Section \ref{qualitative}.

\subsection{Quantitative Results}
\label{quantitative}
We compared the performance of the proposed ScreenTK by comparing it to the SOTA screenshots method, calculating the capture rate of time-killing actions during the explicit and implicit tests. The ground truth of time-killing instances were manually counted by the first author from the sequences of one-second-interval screenshots that were extracted from the screen recordings of the participants' phone usage. For the explicit and implicit tests, a time-killing instance started when a participant clicked an URL in the story or article and ended when the participant returned back to the original reading. We observed that the baseline method of taking screenshots every 5 seconds missed many time-killing instances (57\% for explicit time-killing and 38\% for implicit time-killing). On the other hand, ScreenTK captured almost all time-killing events (except for one instance due to a screen text sensor unresponsive issue). This result suggests that the proposed ScreenTK is significantly more effective in capturing time-killing behaviors compared to the traditional screenshot method.

% using screenshots of every 5 seconds might not be useful to study the dynamic user-phone interactions.

\begin{figure}[t]
    \centering
    \includegraphics[width=\columnwidth]{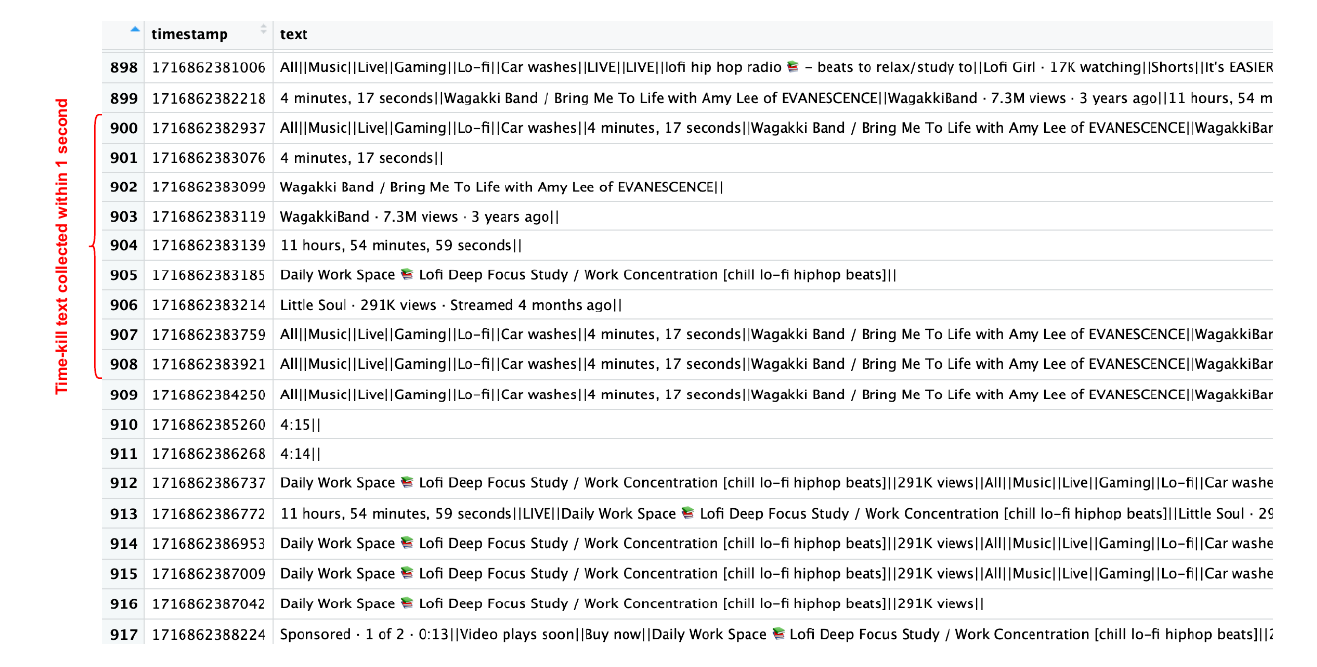} %width=10cm
    \caption{Examples of collected screen text. timestamp: the Unix timestamp at which the event occurred; text: text on the current screen.}
    \label{fig:gap2}
\end{figure}

\subsection{Qualitative Results}
\label{qualitative}
We observed that ScreenTK collected more fine-grained information about time-killing moments when compared to screenshots. As illustrated in Figure \ref{fig:gap1}, the top text box contains the screen-text records about the starting point of the time-killing moment ("Click here.") and the actual content viewed by the participant (i.e., an image file); the bottom text box shows the process of the time-killing action: 1) clicked citation link "[12]", 2) redirected to the URL, and 3) visited the animation. Also, in Figure~\ref{fig:gap2}, we observe that ScreenTK is capable to capture time-killing moment even within in a second period. Specifically, Figure 3 indicates that the participant viewed a music video by the Wagakki Band, featuring a cover of the song "Bring Me to Life" with guest vocals by Amy Lee of Evanescence. Additionally, the participant continued browsing other music videos, evidenced by text like "Daily Work Space Lofi Deep Focus Study." These detailed records highlight ScreenTK's ability to capture user activities with precision, providing a comprehensive view of time-killing behaviors.

\section{Conclusion and Future Work}
\label{conclusion}
In this work, we propose a novel framework called ScreenTK for time-killing detection using continuous screen text and on-device LLMs. Our analysis of experimental results demonstrates that the proposed ScreenTK framework is capable of consistently recording time-killing moments. Compared with screenshot-based methods, ScreenTK provides more comprehensive information about time-killing behavior. 

We noticed that time-killing moments were sometimes related to app-switching behavior that can be monitored by the app sensor. However, the app sensor is not sufficient for time-killing detection. For instance, if a user switches from reading a scientific article to an interesting story on the same website, the app sensor cannot determine the change in attention from educational to time-killing content. In contrast, sentimental analysis of screen text can identify this transition. We found that learning material (e.g., scientific writings) has a more neutral tone and cohesive text compared to unofficial reading. Contents in social media and entertainment are more polarized and less consistent in logic, often containing short, conversational phrases. These findings suggest that screen text can be used to recognize contextual features of time-killing moments.

In conclusion, our study highlights the potential of utilizing screen text data for time-killing detection. By combining app sensor data with screen text analysis, we believe that more accurate time-killing detection can be achieved. For future work, we aim to explore this integration to enable personalized interventions for unwanted phone usage, empowering users with better self-control over their digital life on smartphones.

% For future work, we aims to utilize screen text with on-device LLMs for the time-killing detection, in order to protect users' privacy and data security. We are also interested in exploring automated phone-use annotation of screenshots by referring to the corresponding screen text. In other words, the screen text might be useful to label and classify screenshots.

\begin{acks}
This work is partially funded by NHMRC grants 1170937 and 2004316, AUSMURI grant 13203896, and the CISCO grant 2021-2327463696.
\end{acks}

%---------------------------------------
\bibliographystyle{sections/ACM-Reference-Format}
\bibliography{sections/ref}
\end{document}